# Enhanced magnetocaloric effect from Cr substitution in Ising lanthanide gallium garnets $Ln_3CrGa_4O_{12}$ ($Ln$ = Tb, Dy, Ho)

*Paromita Mukherjee\*, Siân E. Dutton\**
Cavendish Laboratory, University of Cambridge, JJ Thomson Avenue, Cambridge CB3 0HE, United Kingdom
\*E-mail: pm545@cam.ac.uk, sed33@cam.ac.uk

**Abstract:**

A detailed study on the crystal structure and bulk magnetic properties of Cr substituted Ising type lanthanide gallium garnets $Ln_3CrGa_4O_{12}$ ($Ln$ = Tb, Dy, Ho) has been carried out using room temperature powder X-Ray and neutron diffraction, magnetic susceptibility, isothermal magnetisation and heat capacity measurements. The magnetocaloric effect (MCE) in $Ln_3CrGa_4O_{12}$ is compared to that of $Ln_3Ga_5O_{12}$. In lower magnetic fields attainable by a permanent magnet (≤2 T), Cr substitution greatly enhances the MCE by 20% for $Ln$ = Dy and 120% for $Ln$ = Ho compared to the unsubstituted $Ln_3Ga_5O_{12}$. This is likely due to changes in the magnetic ground state as Cr substitution also significantly reduces the frustration in the magnetic lattice for the Ising type $Ln_3Ga_5O_{12}$.

1. Introduction

Many areas of fundamental and applied scientific research including spintronics and quantum computing as well as devices such as magnetic resonance imaging (MRI) scanners and low temperature sensors (such as those in space detectors) require cooling to low temperatures. This is usually achieved using liquid cryogens such as liquid nitrogen $T > 80$ K, liquid helium ($^4$He) for $T > 2$ K or a mixture of $^3$He and $^4$He for cooling down to 20 mK. However, the increasing scarcity of helium (the $^3$He isotope is even less abundant) and rising costs means that alternatives to cryogens must be explored. One such alternative is solid state magnetic cooling using adiabatic demagnetisation refrigerators (ADRs) which are based on the principle of the magnetocaloric effect (MCE) in magnetic materials. Here the cooling limit is determined by the magnetic ordering temperature, $T_0$, of the material.[1–4]

ADRs using dilute paramagnetic salts are used to cool down to temperatures of few mK.[5–7] However, the poor chemical stability of these materials make them less viable for widespread practical applications. A different approach is to use ceramic materials which have geometrically frustrated magnetic lattices where the geometry of the magnetic lattice prevents all the nearest-neighbour magnetic interactions from being satisfied simultaneously.[8] This suppresses or in some cases, completely inhibits magnetic long range ordering. Geometrically frustrated magnets (GFMs) typically show ordering features at $T_0 \sim \theta_{CW}/10$ where $\theta_{CW}$ is the Curie-Weiss temperature, thereby suppressing the ordering temperature.[9] In complex lanthanide oxides, the highly localised *4f* orbitals have weak magnetic interactions, i.e. $\theta_{CW}$ is small and so when the magnetic lattice is frustrated, ordering is suppressed to even lower *T*. The theoretical magnetic entropy that can be extracted is much higher than in transition metal compounds. GFMs with $Ln^{3+}$ ions are therefore ideal candidates for sub 20 K magnetocaloric materials (MCMs). Another advantage is that the lanthanides are chemically very similar but their magnetic properties vary widely. This allows tuning of the properties for optimisation of the MCE.[10–12]

The lanthanide gallium garnets, $Ln_3Ga_5O_{12}$, are a family of materials, where the magnetic $Ln^{3+}$ spins form two interpenetrating networks of ten membered-rings of corner-sharing triangles leading to a high degree of geometrical frustration.[13] Of these, gadolinium gallium garnet (GGG), which shows no long range ordering down to 25 mK, has been established as a MCM for magnetic refrigeration in the liquid helium temperature regime. The absence of long range ordering, high density of magnetic ions, chemical stability and lack of single ion anisotropy (*L*=0 for $Gd^{3+}$) allowing for the full magnetic entropy ($R\ln[2J+1]$ = 17.29 J K$^{-1}$ mol$_{Gd}^{-1}$) to be extracted in high magnetic fields makes it an ideal MCM for T<20 K.[14–16] In recent years, a number of Gd containing MCMs with better performance at 2 K have been reported[17–20] but GGG continues to be used and serves as the benchmark for MCMs in this temperature regime. However, for all the Gd based magnetocalorics, the change in magnetic entropy is maximised in fields of 5 T or higher. Such high magnetic fields can only be produced using a superconducting magnet which again requires cooling using cryogens. For more practical applications, we need to focus on developing materials with high MCE in fields ≤ 2 T, attainable by a permanent magnet. This has been discussed in a recent study on $Tb(HCO_2)_3$ where the MCE is significantly higher than $Gd(HCO_2)_3$ at higher temperatures and lower fields as the $Tb^{3+}$ have Ising-like spins contrasted with the Heisenberg nature of the $Gd^{3+}$ spins.[21] For $Ln_3Ga_5O_{12}$ the MCE in fields ≤ 2 T is expected to be maximised for the $Ln^{3+}$ having Ising-like spins, such as $Dy_3Ga_5O_{12}$ (DGG), $Ho_3Ga_5O_{12}$ (HoGG) and $Tb_3Ga_5O_{12}$ (TbGG).[22–24] DGG is a more efficient MCM than GGG at fields ≤ 2 T.[25] There have been no detailed studies on the MCE in $Tb_3Ga_5O_{12}$ and $Ho_3Ga_5O_{12}$ but the Ising nature of the spins could lead to large changes in the magnetic entropy at low fields.

Much remains to be explored about the optimisation of the MCE in the $Ln_3Ga_5O_{12}$ family. One approach is to maximise the MCE by chemical substitution. Studies in GGG substituting the magnetic $Gd^{3+}$ site with $Tb^{3+}$ or $Dy^{3+}$ [25–27] and the nonmagnetic $Ga^{3+}$ site with $Al^{3+}$ or $Fe^{3+}$ partially or completely[28–31] have shown to have a measurable impact on the MCE. There is a lot of potential for further research on studying the effect of chemical substitution on the MCE in the different $Ln_3Ga_5O_{12}$.

In this paper, we report on the synthesis, characterisation, bulk magnetic properties and MCE in terbium, dysprosium and holmium gallium garnets substituted with chromium, $Ln_3CrGa_4O_{12}$ ($Ln$ = Tb, Dy, Ho) and compare them with $Ln_3Ga_5O_{12}$. The change in magnetic entropy in TbGG in a field of 2 T is of comparable magnitude to DGG, albeit slightly smaller, while that of HoGG is almost half that of DGG. Cr substitution has a dramatic impact on the magnetic properties with increased transition temperatures and enhanced MCE in all Cr containing samples. Most significantly, the change in magnetic entropy in $Ho_3CrGa_4O_{12}$ is more than twice that of $Ho_3Ga_5O_{12}$ in all measured temperatures and magnetic fields.

## 2. Results

### 2.1. Structural Characterisation

PXRD indicated the formation of phase pure $Ln_3CrGa_4O_{12}$ ($Ln$ = Tb, Dy, Ho). Attempts to synthesise $Ln_3Cr_xGa_{5-x}O_{12}$ ($Ln$ = Tb, Dy, Ho) with $x > 1$, resulted in the formation of $LnCrO_3$ ($Ln$ = Tb, Dy, Cr) perovskite impurities. We conclude that only partial substitution of Ga on the octahedral site with Cr (maximum 1:1 ratio) is possible using this synthetic route. The cubic $Ia\bar{3}d$ structure of $Ln_3Ga_5O_{12}$ ($Ln$ = Tb, Dy, Ho) is preserved on Cr substitution (Figure 1a). In the cubic $Ln_3Ga_5O_{12}$ garnet structure there are three distinct cation sites – dodecahedral occupied by $Ln^{3+}$, octahedral occupied by $Ga^{3+}$ and tetrahedral also occupied by $Ga^{3+}$. The connectivity of magnetic $Ln^{3+}$ ions is shown in Figure 1b.

Combined PXRD + PND structural refinements were carried out for $Ln_3CrGa_4O_{12}$ ($Ln$ = Tb, Ho). For $Dy_3CrGa_4O_{12}$ and $Ln_3Ga_5O_{12}$ ($Ln$ = Tb, Dy, Ho), the structural refinements were carried out using only PXRD. The combined room temperature PXRD + PND Rietveld refinement for $Ho_3CrGa_4O_{12}$ is shown in Figure 2. The crystallographic parameters obtained from the fits for $Ln_3CrGa_4O_{12}$ ($Ln$ = Tb, Dy, Ho) are given in Table 1 and for $Ln_3Ga_5O_{12}$ ($Ln$ = Tb, Dy, Ho) in Table S1. Very little change in lattice parameter is observed on $Cr^{3+}$ substitution. This is expected given the similar size of $Cr^{3+}$ and $Ga^{3+}$ ions. The difference in the neutron scattering factor for Cr ($b_{Cr}$ = 3.635 fm) and Ga ($b_{Ga}$= 7.288 fm)[32] allows for the position of $Cr^{3+}$ to be determined. For $Ln$ = Tb and Ho, $Cr^{3+}$ is found to exclusively occupy the octahedral site, as would be expected from crystal electric field (CEF) considerations. Therefore, it was also assumed that Cr exclusively occupies the octahedral site in $DyCrGa_4O_{12}$. The refined composition for $Ln$ = Tb, Ho was determined to be the same as the nominal composition within error. The composition for $Ln$ = Dy was fixed at the nominal composition as PXRD is not sensitive enough to refine the Cr/Ga occupancy. We will use the nominal composition in all further discussions.

On Cr substitution, the changes in the $Ln$-O, $Ln$-$Ln$, Cr/Ga1-O and Ga2-O bond lengths are small and almost all within error for $Ln$ = Tb, Dy, Ho (Table S2). Therefore, no significant change in $Ln^{3+}$ single-ion anisotropy is expected on Cr substitution. The resultant change in the dipolar interaction ($D \propto 1/r_{Ln-Ln}^3$) between adjacent $Ln^{3+}$ ions on Cr substitution is also small, less than 0.1% for all samples.

## 2.2. Magnetic Measurements

The Zero Field Cooled (ZFC) magnetic susceptibility, $\chi(T)$, of $Ln_3CrGa_4O_{12}$ ($Ln$ = Tb, Dy, Ho) measured in a field of 100 Oe from 2- 300 K is shown in Figure 3 and in Figure S1 for $Ln_3Ga_5O_{12}$ ($Ln$ = Tb, Dy, Ho). No long-range magnetic ordering is observed down to 2 K for any sample. Above $T > 100$ K, the inverse susceptibility $\chi^{-1}$ is linear and fits to the Curie-Weiss law for the Cr substituted garnets are summarised in Table 2. The negative values of $\theta_{CW}$ indicate antiferromagnetic correlations. However, these values do not account for low-lying crystal-field effects which may significantly alter the value of $\theta_{CW}$. The experimental values of magnetic moment obtained from fitting to the Curie-Weiss law are consistent with the theoretical values, $\mu_{eff}^2 = 3\mu_{Ln}^2 + \mu_{Cr}^2$.

The isothermal magnetisation $M(H)$ for $Ln_3CrGa_4O_{12}$ ($Ln$ = Tb, Dy, Ho) measured at different temperatures is shown in Figure 4 and the maximum values of magnetisation/f.u.$_{Ln}$, $M_{max}$, are collated in Table 2. Figure S2 shows the $M(H)$ curves for $Ln_3Ga_5O_{12}$ ($Ln$ = Tb, Dy, Ho). The $M(H)$ increases rapidly at fields $\leq$ 2 T and increases more gradually in higher fields. None of the $M(H)$ curves attain saturation in the experimentally limiting field of 9 T. The observed $M_{max}$ for all samples is much lower than the theoretical saturation value for a Heisenberg system, $M_{sat} = 3g_J J + g_S S$. However, for the Ising-like Tb$^{3+}$, Dy$^{3+}$ and Ho$^{3+}$ spins in $Ln_3Ga_5O_{12}$,[22–24] the saturation value for the magnetic $Ln^{3+}$ is expected to be close to $3g_J J/2$.

Our $M_{max}$ values are in agreement with this and so we propose that the Ising nature of the $Ln^{3+}$ ($Ln$ = Tb, Dy, Ho) is retained on Cr substitution in these garnets. Thus the MCE for the Cr substituted garnets is also expected to be optimized in fields up to 2 T. The contribution of $Cr^{3+}$ spins to the net magnetisation is very small compared to $Ln^{3+}$ and so we cannot definitively comment on their nature. However, $Cr^{3+}$ ($d^3$) spins are likely to have Heisenberg nature[33] and the $M_{max}$ values are consistent with this.

### 2.3. Heat Capacity Measurements

The magnetic heat capacity, $C_{mag}$, as a function of temperature and field for $Ln_3CrGa_4O_{12}$ ($Ln$ = Tb, Dy, Ho) is shown in Figure 5. Inset shows $C_{mag}/T$ in zero field measured down to 0.5 K. The lattice contribution was subtracted using a Debye model[34] with $\theta_D$ = 360 K for $Tb_3CrGa_4O_{12}$, $\theta_D$ = 340 K for $Dy_3CrGa_4O_{12}$ and $\theta_D$ = 330 K for $Ho_3CrGa_4O_{12}$. The nuclear Schottky anomaly for $Ho_3CrGa_4O_{12}$ was subtracted using a model for the hyperfine interactions for $HoCrO_3$.[35] In zero field, $Tb_3CrGa_4O_{12}$, $Dy_3CrGa_4O_{12}$ and $Ho_3CrGa_4O_{12}$ show magnetic ordering features at 1.72 K, 1.75 K and 1.55 K respectively. In higher fields the ordering transition shifts to higher temperatures and is broadened. At 9 T, the transition manifests as a very broad feature at ~5, 10 and 12 K for $Ln_3CrGa_4O_{12}$ ($Ln$= Tb, Dy, Ho) respectively.

TbGG, DGG and HoGG are reported to order at 0.25 K, 0.373 K and 0.19 K respectively.[24,36] Cr substitution significantly enhances the transition temperature for these Ising type lanthanide gallium garnets. $T_0$ is increased from 0.25 K to 1.72 K for $Ln$ = Tb, from 0.373 K to 1.75 K for $Ln$ = Dy and from 0.19 K to 1.55 K for $Ln$ = Ho. The frustration index, $f$, of the magnetic lattice, defined by $f = \theta_{CW}/T_0$ is reduced on Cr substitution (Table 2). In the most extreme case, for $Dy_3CrGa_4O_{12}$, the frustration index is only slightly higher than would be expected for an antiferromagnet and is significantly reduced from $f \sim 23$ for $Dy_3Ga_5O_{12}$. The origin of the changes in the magnetic frustration is not clear. It has been reported that increased single-ion anisotropy in the $Ln_3Al_5O_{12}$ garnets increases $T_0$.[37] However, the lack of any significant changes in the *Ln*-O environment would suggest that this is not the case here. Further experiments using PND are required to determine a) the nature of magnetic ordering in $Ln_3CrGa_4O_{12}$ (*Ln*= Tb, Dy, Ho) b) whether there are any differences in the magnetic ground state for the different *Ln* compared to the unsubstituted gallium garnets.

**2.4. Magnetocaloric Effect**

The magnetocaloric effect (MCE) can be characterised by the change in magnetic entropy, $\Delta S_m$, per mole which can be calculated from the *M(H)* curves using Maxwell's thermodynamic relation:[38]

$$\Delta S_m = \int_{H_{initial}}^{H_{final}} \left(\frac{\partial M}{\partial T}\right)_H dH$$

Contour plots for $\Delta S_m$ per mole as a function of temperature *T* and magnetic field $\mu_0 H$ for $Ln_3CrGa_4O_{12}$ (*Ln* = Tb, Dy, Ho) are given in Figure 6. The MCE is compared to unsubstituted $Ln_3Ga_5O_{12}$ (*Ln* = Tb, Dy, Ho) in a field of 2 T in Figure 7. Insets show the variation of $\Delta S_m$ per mole in field at 2 K. Figure 7 shows that DGG has a higher $\Delta S_m$ than TbGG and HoGG. $\Delta S_m$ for TbGG at 2 K, 2 T (8.66 J K$^{-1}$ mol$^{-1}$) is lower than DGG (11.32 J K$^{-1}$ mol$^{-1}$); however for HoGG, the change in $\Delta S_m$ is about half that for DGG. These differences could be due to differences in the nature of the magnetic ordering in these garnets.

On Cr substitution, $Dy_3CrGa_4O_{12}$ has the largest change in magnetic entropy over the entire temperature and field range among the three Cr substituted garnets. The contour plots (Figure 6) also show the onset of field-induced transitions at fields > 2 T for all three Cr substituted samples. The most pronounced field induced transition is observed in $DyCrGa_4O_{12}$ at ~ 5 K and 5 T, further experiments are required to investigate the nature of the observed transitions. The differences in the effect of Cr substitution on the MCE for the different Ising-type $Ln^{3+}$ is striking. For *Ln* = Tb, the difference in the MCE on Cr substitution is minimal at low fields, however an increase in $\Delta S_m$ is observed at fields above 2.5 T. Whereas for *Ln* = Dy, there is a 20% increase in $\Delta S_m$ in a field of 2 T. The most dramatic increase in $\Delta S_m$ is seen for $Ln$ = Ho where $\Delta S_m$ shows ~120% increase for $Ho_3CrGa_4O_{12}$ compared to $Ho_3Ga_5O_{12}$ in a field of 2 T. Thus, substitution with Cr significantly enhances the MCE in DGG and HoGG (especially $Ln$ = Ho) in magnetic fields ≤ 2 T and temperatures below 10 K. The origin of the impact of Cr substitution on $\Delta S_m$ is likely due to changes in the magnetic ordering and nature of the magnetic ground state, indicated by the dramatic change in the magnetic frustration and enhancement of the ordering temperature for $Ln_3CrGa_4O_{12}$. For a MCM, the maximum change in the magnetic entropy is obtained around the ordering temperature, $T_0$. The minimum temperature for our $\Delta S_m$ (T) calculations, 2 K, is very close to $T_0$ for $Ln_3CrGa_4O_{12}$ (Table 2) and so, an enhancement in $\Delta S_m$ is observed.

The values of $\Delta S_m$ for $Ln_3CrGa_4O_{12}$ ($Ln$ = Dy, Ho) in gravimetric units are comparable to the maximum values reported for other rare-earth transition metal systems like $Ln$CrO$_4$ and $Ln$VO$_4$ ($Ln$ = Dy, Ho).[39–41] However for these systems, $\Delta S_m$ is maximized at higher temperatures, $T$ > 20 K, restricting their use for cooling in the liquid helium temperature regime. $Ln_3CrGa_4O_{12}$ ($Ln$ = Dy, Ho), however, can be used as effective MCMs for $T \geq 2$ K in fields up to 2 T. Further $Ln_3CrGa_4O_{12}$ and $Ln_3Ga_5O_{12}$ ($Ln$ = Dy, Ho) could be used to develop potential graded magnetocalorics so that the cooling limit is further reduced to $T \geq 0.4$ K ($T_0$ for the $Ln_3Ga_5O_{12}$).

## 3. Conclusion

We have prepared powder samples of $Ln_3CrGa_4O_{12}$ ($Ln$ = Tb, Dy, Ho) and carried out a detailed investigation of the crystal structure and bulk magnetic properties. The MCE has been calculated and compared to unsubstituted $Ln_3Ga_5O_{12}$ ($Ln$ = Tb, Dy, Ho). It is seen that in lower magnetic fields, $\mu_0 H \leq 2$ T, Cr substitution greatly enhances the MCE in Ising type lanthanide gallium garnets - by 20% for $Ln$ = Dy and 120% for $Ln$ = Ho in a field of 2 T. These materials are viable MCMs in the liquid helium temperature regime ($T \geq 2$ K). The enhancement in MCE is postulated to be due to the changes in magnetic ordering as Cr substitution also significantly reduces the magnetic frustration and enhances the transition temperature in $Ln_3Ga_5O_{12}$ ($Ln$ = Tb, Dy, Ho).

## 4. Experimental Section

*Sample preparation*: Powder samples of $Ln_3CrGa_4O_{12}$ ($Ln$ = Tb, Dy, Ho) were prepared using a solid-state synthesis by mixing stoichiometric amounts of $Tb_4O_7$ (99.999% purity, Alfa Aesar) or $Dy_2O_3$ (99.999% purity, Alfa Aesar) or $Ho_2O_3$ (99.999% purity, Alfa Aesar), $Ga_2O_3$ (99.99% purity, Alfa Aesar) and $Cr_2O_3$ (99.99% purity, Alfa Aesar). $Ga_2O_3$ was pre-dried at 500 °C prior to weighing out to ensure accurate chemical composition. The powders were intimately mixed and pressed into pellets which were heated between 1200 – 1400 °C for 48-72 hours with intermittent regrindings. Samples of $Ln_3Ga_5O_{12}$ ($Ln$ = Tb, Dy, Ho) were prepared in a similar fashion except heat treatments were only carried out at 1200 °C as described elsewhere.[31]

*Structural Characterisation:* Powder X-Ray diffraction (PXRD) was used to confirm the formation of phase pure products. Initially short scans were collected over $10° \leq 2\theta \leq 60°$ ($\Delta 2\theta = 0.015°$) using a Panalytical Empyrean X-Ray diffractometer (Cu Kα radiation, $\lambda$ = 1.540 Å). Longer scans over a wider angular range $10° \leq 2\theta \leq 90°$ ($\Delta 2\theta = 0.008°$) were collected for quantitative structural analysis. Room temperature (RT) powder neutron diffraction (PND) experiments for structural characterisation were carried out on the D2B diffractometer, Institut Laue-Langevin (ILL), ($\lambda$ = 1.594791(6) Å) at 300 K for $Ln_3CrGa_4O_{12}$ ($Ln$ = Tb, Ho). The structural Rietveld refinement was carried out using the Fullprof suite of programmes.[42] Linear interpolation was used to fit the background and the peak shape was modelled using a pseudo-Voigt function.

*Magnetic Measurements:* Magnetic properties were measured on a Quantum Design Magnetic Properties Measurement System (MPMS) with a Superconducting Quantum Interference

Device (SQUID). The zero-field cooled (ZFC) susceptibility was measured in a field of 100 Oe in the temperature range 2 – 300 K for $Ln_3CrGa_4O_{12}$ and $Ln_3Ga_5O_{12}$ ($Ln$ = Tb, Dy, Ho). In a small field of 100 Oe the linear approximation for molar susceptibility $\chi(T) \sim M/H$ is taken to be valid where $M$ is the magnetisation and $H$ is the magnetic field. Isothermal magnetisation measurements in the field range 0 – 9 T for selected temperatures were carried out using the ACMS (AC measurement system) option on the Quantum Design Physical Properties Measurement System (PPMS).

*Heat Capacity Measurements:* Heat capacity measurements for $Ln_3CrGa_4O_{12}$ ($Ln$ = Tb, Dy, Ho) were performed using a Quantum Design PPMS in the temperature range 1.8 – 30 K in fields 0 - 9 T. Equal amounts of sample and silver powder (99.99% Alfa Aesar) were mixed and pressed into pellets which were then used for measurement. To obtain the sample heat capacity, the contribution from silver was subtracted using values from the literature.[43] Additional measurements were made using the He3 option down to 0.5 K in zero field. Further details of the crystal structure investigation(s) may be obtained from the Fachinformationszentrum Karlsruhe, 76344 Eggenstein-Leopoldshafen (Germany), on quoting the depository number CSD - 432904 for $Tb_3CrGa_4O_{12}$, CSD - 432905 for $Dy_3CrGa_4O_{12}$ and CSD - 432906 for $Ho_3CrGa_4O_{12}$.

**Acknowledgements**

We thank Dr. Emmanuelle Suard, D2B Instrument Responsible, for her support as local contact in carrying out the neutron diffraction experiments on D2B, ILL and for valuable feedback. We thank H.F.J. Glass for his support during the neutron diffraction experiments on D2B, ILL and for useful discussions.

We acknowledge funding support from the Winton Programme for the Physics of Sustainability. Magnetic measurements were carried out using the Advanced Materials Characterisation Suite, funded by EPSRC Strategic Equipment Grant EP1M00052411.

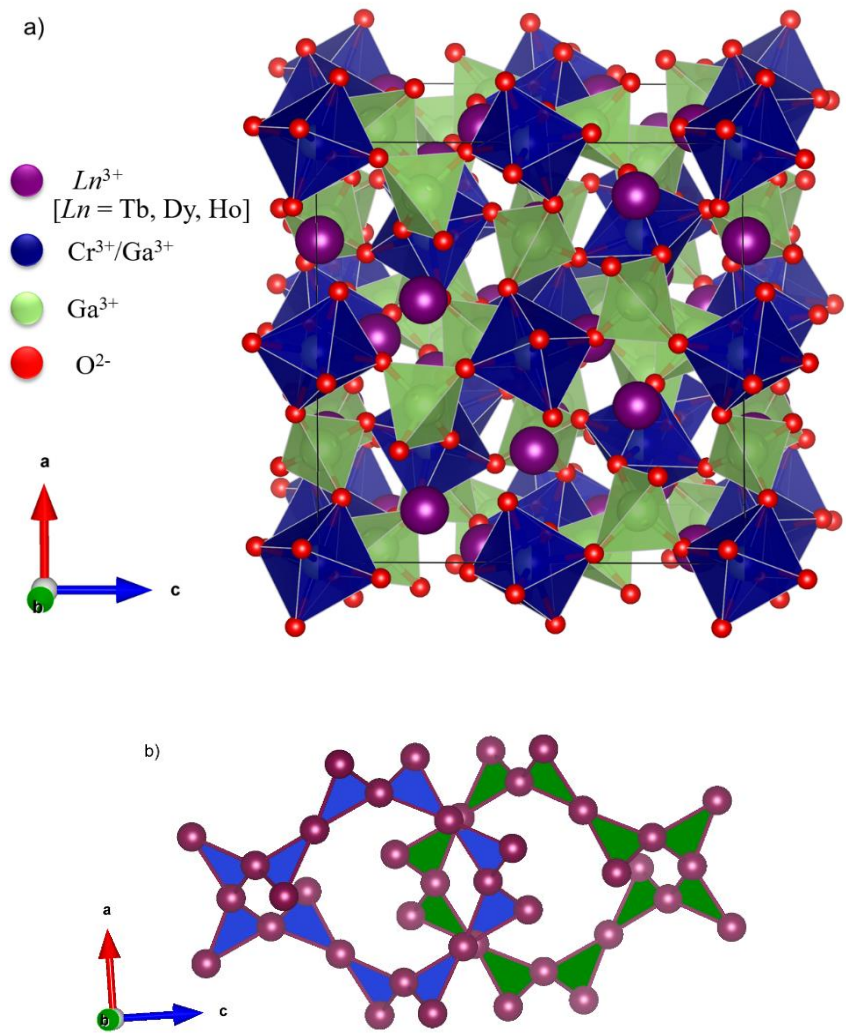

**Figure 1.** a) Crystal structure of $Ln_3CrGa_4O_{12}$ ($Ln$ = Tb, Dy, Ho): $Ln^{3+}$ ($Ln$ = Tb, Dy, Ho) occupies the dodecahedral site, $Cr^{3+}/Ga^{3+}$ are disordered over the octahedral site, $Ga^{3+}$ occupies the tetrahedral site and $O^{2-}$ occupies the general ($x$, $y$, $z$) position b) Connectivity of magnetic $Ln^{3+}$ ($Ln$ = Tb, Dy, Ho) lattice: The $Ln^{3+}$ lie at the vertices of corner-sharing equilateral triangles in two interpenetrating ten-membered rings, forming a highly frustrated three-dimensional network.

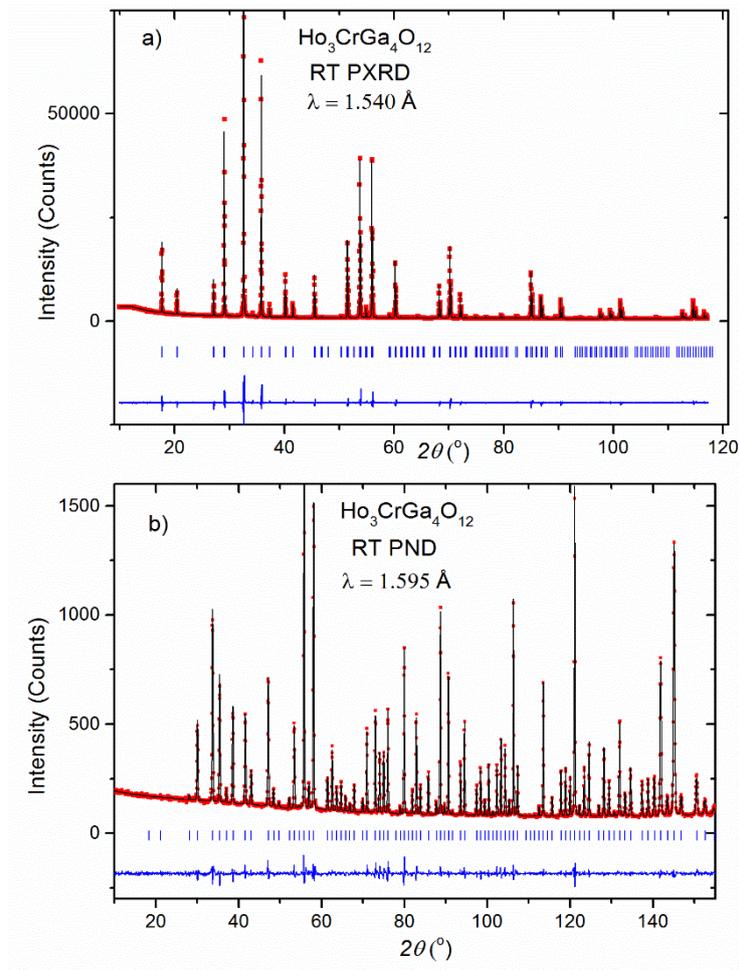

**Figure 2.** Room temperature a) PXRD b) PND pattern for Ho$_3$CrGa$_4$O$_{12}$: Experimental data (∎), Modelled data (-), Difference pattern (-), Bragg positions (│).

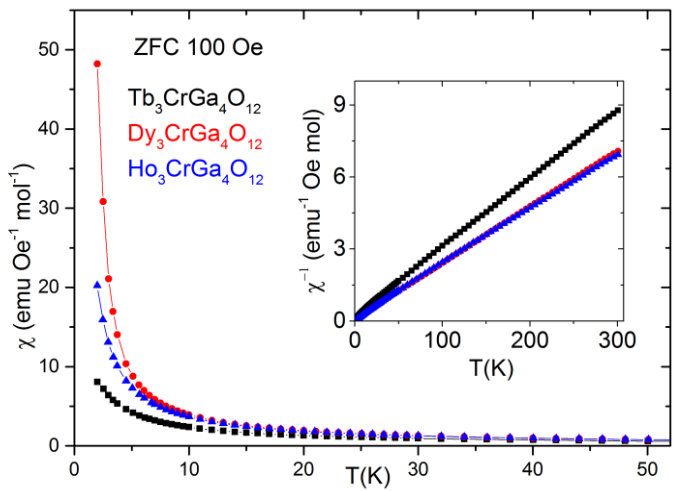

**Figure 3.** ZFC molar susceptibility $\chi(T)$ for $Ln_3$CrGa$_4$O$_{12}$ ($Ln$ = Tb, Dy, Ho) measured from 2-300 K in a field of 100 Oe; inset: inverse molar susceptibility $\chi^{-1}(T)$.

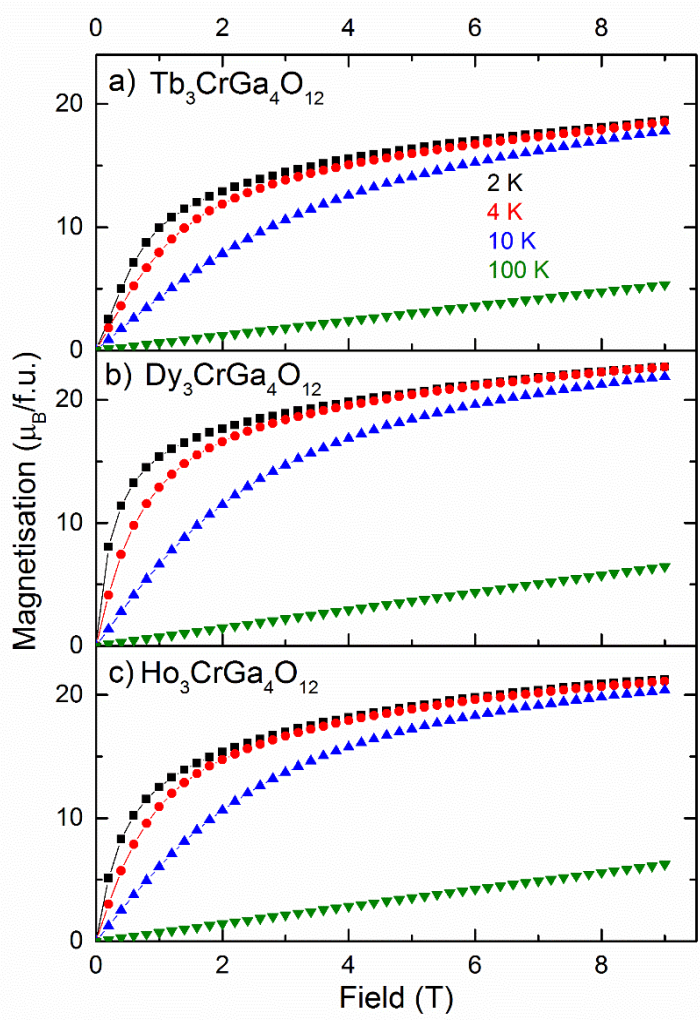

**Figure 4.** Isothermal *M(H)* curves measured from 0 – 9 T at select temperatures for *Ln*$_3$CrGa$_4$O$_{12}$ – a) *Ln* = Tb  b) *Ln* = Dy  c) *Ln* = Ho.

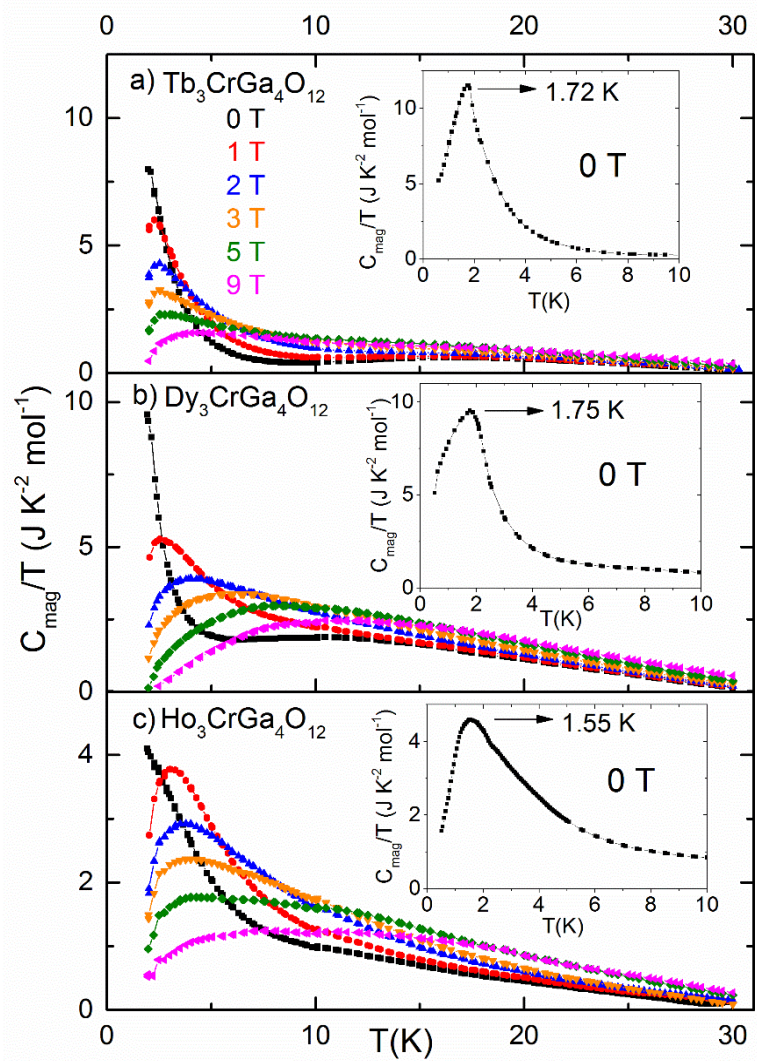

**Figure 5.** Heat capacity in different magnetic fields, inset: zero field heat capacity measured down to 0.5 K for $Ln_3CrGa_4O_{12}$ – a) $Ln$ = Tb b) $Ln$ = Dy c) $Ln$ = Ho.

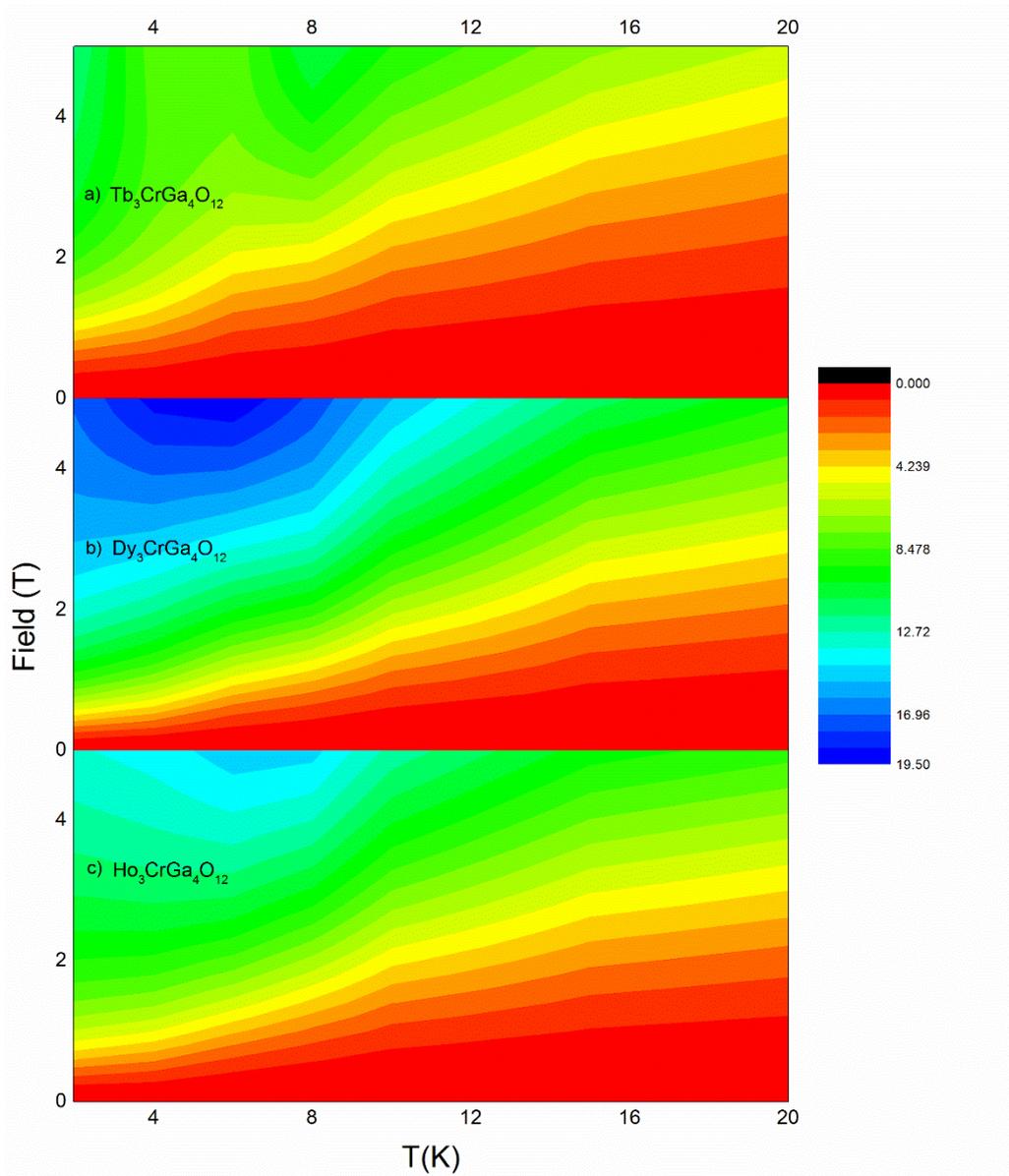

**Figure 6.** Contour plots for $\Delta S_m$ per mole as a function of temperature and external magnetic field for $Ln_3CrGa_4O_{12}$ – a) $Ln$ = Tb b) $Ln$ = Dy c) $Ln$ = Ho.

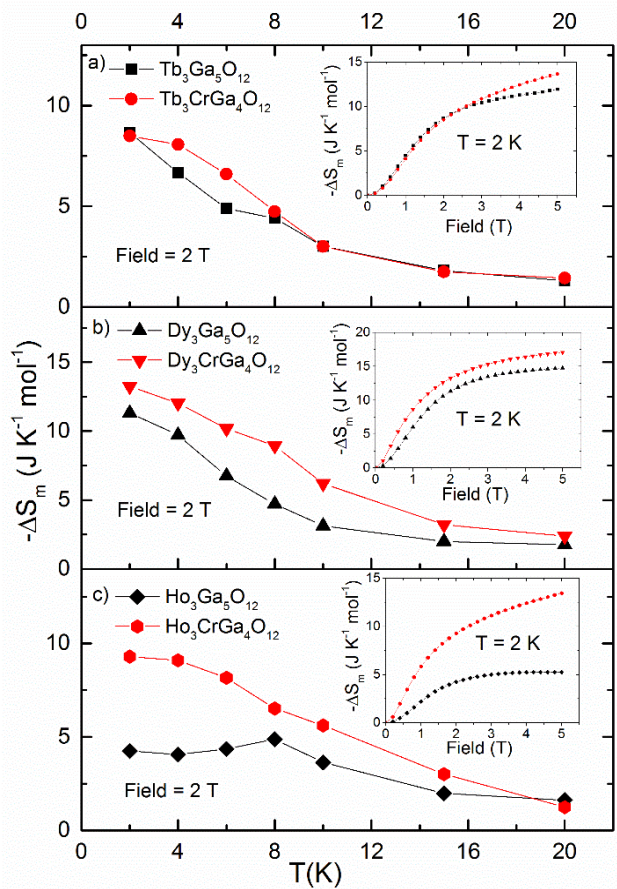

**Figure 7.** $\Delta S_m$ per mole as a function of temperature for $Ln_3CrGa_4O_{12}$ compared to $Ln_3Ga_5O_{12}$ ($Ln$ = Tb, Dy, Ho) in a field of 2 T, inset: field dependence of $\Delta S_m$ per mole at $T$ = 2 K – a) $Ln$ = Tb b) $Ln$ = Dy c) $Ln$ = Ho.

**Table 1.** Crystallographic parameters for $Ln_3CrGa_4O_{12}$ ($Ln$ =Tb, Dy, Ho). The structural Rietveld refinements were carried out in the $Ia\bar{3}d$ space group, with $Ln$ ($Ln$ = Tb, Dy, Ho) on the dodecahedral 24$c$ (0, 0.25, 0.125) site, Cr/Ga1 on the octahedral 16$a$ (0, 0, 0) site, Ga2 on the tetrahedral 24$d$ (0, 0.25, 0.375) site and O on the 96$h$ ($x$, $y$, $z$) site.

|  |  | $Tb_3CrGa_4O_{12}$ | $Dy_3CrGa_4O_{12}$[a] | $Ho_3CrGa_4O_{12}$ |
|---|---|---|---|---|
|  | $a$ (Å) | 12.34563(6) | 12.30144 (21) | 12.28390 (7) |
| $Ln$ | $B_{iso}$ (Å$^2$) | 0.182 (19) | 0.5 | 0.041 (17) |
| Cr/Ga1 | Frac Cr | 0.51(1) | 0.50[b] | 0.50(1) |
|  | $B_{iso}$ (Å$^2$) | 0.21 (4) | 0.5 | 0.16 (4) |
| Ga2 | $B_{iso}$ (Å$^2$) | 0.281 (20) | 0.5 | 0.270 (18) |
| O | $x$ | -0.02845 (6) | -0.0287 (3) | -0.02784 (6) |
|  | $y$ | 0.05477 (8) | 0.0534 (3) | 0.05559 (7) |
|  | $z$ | 0.14987 (7) | 0.1501 (4) | 0.15006 (7) |
|  | $B_{iso}$ (Å$^2$) | 0.349 (13) | 0.5 | 0.347 (12) |
|  | $\chi 2$ | 2.84 | 2.33 | 3.02 |

[a] Parameters refined from PXRD only, $B_{iso}$ fixed for all site positions; [b] Nominal composition assumed.

**Table 2.** Bulk magnetisation parameters for $Ln_3CrGa_4O_{12}$ ($Ln$ = Tb, Dy, Ho); $T_0$ is determined from zero field heat capacity data

| Sample | $T_0$ (K) | $\theta_{CW}$ (K) | f = $\|\theta_{CW}\|/T_0$ | Theoretical $\mu_{eff}$ ($\mu_B$) | Experimental $\mu_{eff}$ ($\mu_B$) | Theoretical $M_{sat}$ = $g_J J$ ($\mu_B$/f.u.) | M at T = 2 K, H = 9 T ($\mu_B$/f.u.) |
|---|---|---|---|---|---|---|---|
| $Tb_3CrGa_4O_{12}$ | 1.72(5) | -9.7(1.7) | 5.6 | 17.27 | 16.77(8) | 30 | 18.7 |
| $Dy_3CrGa_4O_{12}$ | 1.75(5) | -4.1(1.5) | 2.3 | 18.85 | 18.50(7) | 33 | 22.7 |
| $Ho_3CrGa_4O_{12}$ | 1.55(5) | -9 (3) | 5.8 | 18.78 | 18.80(19) | 33 | 21.2 |

**Supporting Information**

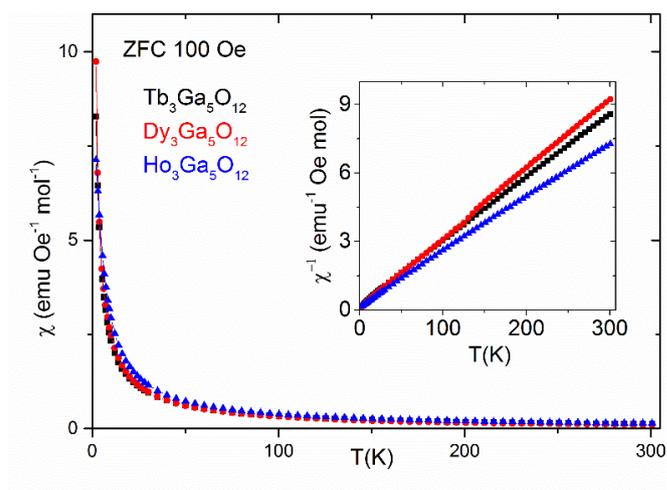

**Figure S1.** ZFC molar susceptibility $\chi(T)$ for $Ln_3Ga_5O_{12}$ ($Ln$ = Tb, Dy, Ho) measured from 2-300 K in a field of 100 Oe; inset: inverse molar susceptibility $\chi^{-1}(T)$

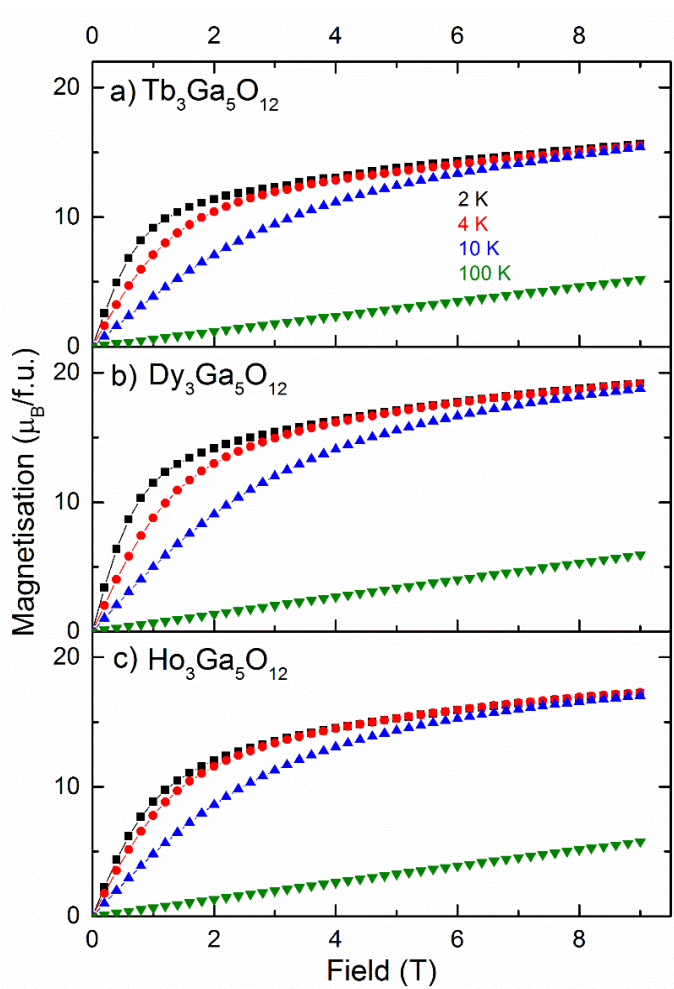

**Figure S2.** Isothermal $M(H)$ curves measured from 0 – 9 T at select temperatures for $Ln_3Ga_5O_{12}$ ($Ln$ = Tb, Dy, Ho)

**Table S1.** Crystallographic parameters for $Ln_3Ga_5O_{12}$ ($Ln$ =Tb, Dy, Ho). The structural Rietveld refinements were carried out in the $Ia\bar{3}d$ space group, with $Ln$ ($Ln$ = Tb, Dy, Ho) on the dodecahedral 24$c$ (0, 0.25, 0.125) site, Ga1 on the octahedral 16$a$ (0, 0, 0) site, Ga2 on the tetrahedral 24$d$ (0, 0.25, 0.375) site and O on the 96$h$ ($x$, $y$, $z$) site. Parameters were refined from PXRD only and $B_{iso}$ was fixed for all site positions.

|     |                   | $Tb_3Ga_5O_{12}$ | $Dy_3Ga_5O_{12}$ | $Ho_3Ga_5O_{12}$ |
|-----|-------------------|------------------|------------------|------------------|
|     | $a$ (Å)           | 12.34191 (4)     | 12.31057 (5)     | 12.28157 (5)     |
| $Ln$ | $B_{iso}$ (Å$^2$) | 0.5              | 0.5              | 0.5              |
| Ga1 | $B_{iso}$ (Å$^2$) | 0.5              | 0.5              | 0.5              |
| Ga2 | $B_{iso}$ (Å$^2$) | 0.5              | 0.5              | 0.5              |
| O   | $x$               | -0.03072 (28)    | -0.02986 (28)    | -0.02976 (23)    |
|     | $y$               | 0.0541 (3)       | 0.0539 (3)       | 0.05150 (28)     |
|     | $z$               | 0.1499 (3)       | 0.1494 (4)       | 0.14942 (29)     |
|     | $B_{iso}$ (Å$^2$) | 0.5              | 0.5              | 0.5              |
|     | $\chi^2$          | 2.92             | 2.59             | 3.60             |

**Table S2**. Selected bond lengths from room temperature powder X-ray diffraction refinements for $Ln_3Ga_5O_{12}$ and $Ln_3CrGa_4O_{12}$ ($Ln$ = Tb, Dy, Ho). Bond lengths were determined from analysis of room temperature PXRD only to enable consistent comparisons.

| $Ln$ | Tb | | Dy | | Ho | |
|---|---|---|---|---|---|---|
|  | $Ln_3Ga_5O_{12}$ | $Ln_3CrGa_4O_{12}$ | $Ln_3Ga_5O_{12}$ | $Ln_3CrGa_4O_{12}$ | $Ln_3Ga_5O_{12}$ | $Ln_3CrGa_4O_{12}$ |
| $Ln$-$Ln$ (Å) | 3.77892 (4) × 4 | 3.77994 (6) × 4 | 3.76933 (5) × 4 | 3.76891 (3) × 4 | 3.76045 (5) × 4 | 3.76117 (7) × 4 |
| $Ln$-O (Å) | 2.380 (5) × 4<br>2.467 (5) × 4 | 2.373 (4) × 4<br>2.462 (4) × 4 | 2.368 (5) × 4<br>2.461 (5) × 4 | 2.350 (5) × 4<br>2.465 (4) × 4 | 2.354 (4) × 4<br>2.483 (3) × 4 | 2.364 (3) × 4<br>2.462 (3) × 4 |
| <$Ln$-O> (Å) | 2.424 | 2.418 | 2.415 | 2.408 | 2.419 | 2.413 |
| Ga1 -O or Cr/Ga1-O (Å) | 2.003 (4) × 6 | 1.992 (3) × 6 | 1.991 (4) × 6 | 1.993 (4) × 6 | 1.975 (3) × 6 | 1.981 (3) × 6 |
| Ga2-O (Å) | 1.824 (5) × 4 | 1.842 (4) × 4 | 1.828 (5) × 4 | 1.830 (4) × 4 | 1.815 (4) × 4 | 1.823 (3) × 4 |